\documentclass[twocolumn]{aastex631}

\newcommand{\msun}{\ensuremath{\textrm{M}_{\odot}}}
\newcommand{\mstar}{\ensuremath{\textrm{M}_{\ast}}}
\newcommand{\lgmstar}{\ensuremath{\log_{10}(\mstar/\msun)}}
\newcommand{\ha}{\ensuremath{\textrm{H}\alpha}}
\newcommand{\dbalmer}{\ensuremath{\textrm{D}_{n}(4000)}}
\newcommand{\ewhd}{\ensuremath{\textrm{EW}(\textrm{H}\delta_{A})}}
\newcommand{\ewha}{\ensuremath{\textrm{EW}(\textrm{H}\alpha)}}
\usepackage{cancel}
\usepackage{ulem}
\usepackage{amsmath}
\begin{document}

\title{Post-starburst galaxies in SDSS-IV MaNGA: Two broad categories of evolutionary pathways}

\correspondingauthor{Zhuo Cheng}
\email{chengz18@mails.tsinghua.edu.cn}

\correspondingauthor{Cheng Li}
\email{cli2015@tsinghua.edu.cn}

\author{Zhuo Cheng}
\affiliation{Department of Astronomy, Tsinghua University, Beijing 100084, China}

\author[0000-0002-8711-8970]{Cheng Li}
\affiliation{Department of Astronomy, Tsinghua University, Beijing 100084, China}

\author[0000-0002-0656-075X]{Niu Li}
\affiliation{Department of Astronomy, Tsinghua University, Beijing 100084, China}

\author[0000-0003-1025-1711]{Renbin Yan}
\affiliation{Department of Physics, The Chinese University of Hong Kong, Sha Tin, NT, Hong Kong, China}

\author[0000-0001-5356-2419]{Houjun Mo}
\affiliation{Department of Astronomy, University of Massachusetts, Amherst MA 01003-9305, USA}

\begin{abstract}
We study the size-mass relation (SMR) and recent star formation history (SFH)
of post-starburst (PSB) galaxies in the local Universe, using spatially
 resolved spectroscopy from the final data release of MaNGA.
 Our sample includes 489 PSB galaxies: 94 cPSB galaxies
with central PSB regions, 85 rPSB galaxies with ring-like PSB regions
and 310 iPSB galaxies with irregular PSB regions.
When compared to control galaxies of similar SFR, redshift and mass,
a similar SMR is found for all types of PSB samples
except the cPSB galaxies which have smaller sizes at
intermediate masses ($9.5\lesssim \log_{10}(\rm M_\ast/M_\odot)\lesssim 10.5$).
The iPSB galaxies in the star-forming sequence (iPSB-SF)
show no/weak gradients in
\dbalmer, \ewhd~and~\ewha, consistent with the global
star-forming status of this type of galaxies, while
the quiescent iPSB (iPSB-Q) sample shows negative gradients in \dbalmer\
and positive gradients in \ewhd,
indicating older stellar populations in the inner regions.
Both cPSB and rPSB samples show positive
gradients in \dbalmer~and negative gradients in \ewhd,
indicating younger stellar populations in the inner regions.
These results imply that the four types of PSB galaxies can be
broadly divided into two distinct categories in terms of
evolutionary pathway:
(1) iPSB-SF and iPSB-Q which have SMRs and SFHs similar to
control galaxies, preferring an inside-out quenching process,
(2) rPSB and cPSB which appear to be different stages of the
same event, likely to follow the outside-in quenching process
driven by disruption events such as mergers that result in a more
compact structure as quenching proceeds.
\end{abstract}


\section{Introduction} \label{sec:intro}

Post-starburst (PSB) galaxies, also known as ``E+A'' or ``K+A'' galaxies, 
are identified by strong Balmer absorption lines but no/weak \ha\ or [O {\sc ii}] 
emission lines in their optical spectrum
\citep[e.g.][]{Dressler1983,Poggianti1999,Goto2005}. 
The lack of strong [O {\sc ii}] or \ha\ emission indicates no or weak 
ongoing star formation, while strong Balmer absorption lines indicate a 
starburst within the past $\sim$1Gyr.
Thus the two spectral features combine to indicate a burst of star 
formation followed by a rapid cessation process of the star formation 
with a timescale of $<1$Gyr. 
The PSB galaxies account for only about 1\% of the galaxy population 
in the local Universe \citep[e.g.][]{Goto2008,Wong2012,Meusinger2017,Pawlik2018}.
Many of them fall in the ``green valley'' of 
color-mass diagrams \citep[e.g.][]{Wong2012}, a 
relatively rare population located in between the two major populations 
of galaxies: the star-forming blue cloud and the quiescent red sequence
\citep[e.g.][]{Strateva2001,Kauffmann2003,Baldry2004,Bell2004,Bundy2005,
	Arnouts2007,Faber2007}. Therefore, PSB galaxies are widely believed
to be transitioning from the blue cloud to the red sequence, 
thus playing an important role in galaxy evolution, 
although the exact fraction of quiescent galaxies that have undergone a
PSB phase remains elusive
\citep[e.g.][]{Tran2003,Tran2004,Yang2008,Wild2009,Vergani2010,
	Swinbank2012,Whitaker2012,Wong2012,Pawlik2016,Yesuf2022}.  

Despite a rich history of studies in the past 
four decades, the physical origin of the fast quenching in PSB galaxies 
is still under debate. In early studies, since these galaxies were preferentially
found in galaxy clusters \citep[e.g.][]{Dressler1983,Poggianti1999,Tran2003},
mechanisms occurring in dense environments such as
perturbations by the cluster tidal field
\citep{Byrd1990} and repeated encounters with companion galaxies 
known as ``harassment'' \citep{Moore1996,Moore1998}
were considered. In more recent studies ram-pressure stripping 
of the cold interstellar medium \citep{Gunn1972} has also been 
considered \citep[e.g.][]{Paccagnella2017,Vulcani2020,Wilkinson2021}. 
However, \cite{Yesuf2022} found that the satellite-quenching mechanisms 
operating in high-density environments cannot account for most of their PSB 
galaxies selected from the Sloan Digital Sky Survey \citep[SDSS;][]{York2000}.

Gas removal driven by active galactic nuclear (AGN) activity has been 
suggested as an alternative mechanism to quickly shut down the star formation 
during the PSB phase
\citep[e.g.][]{Goto2006,Yan2006,Wild2007,Sell2014,Yesuf2014}. 
However, A recent study of Chandra observations by \cite{Lanz2022} 
showed that 
PSB galaxies can contain only low-luminosity AGN, which are unable to 
radiatively drive out the cold gas in low$-z$ PSB galaxies, as detected by 
many studies \citep[e.g.][]{Chang2001,Buyle2006,Zwaan2013,French2015,
	Yesuf2017,Bezanson2022,French2022}.

In fact, it has been well-established that PSB galaxies generally reside in 
the field and loose groups rather than dense
environments based on large spectroscopic surveys, both at low redshift
such as the Las Campanas Redshift Survey \citep[e.g.][]{Zabludoff1996}, 
the 2dF Galaxy Redshift Survey \citep[e.g.][]{Blake2004} and the SDSS \citep[e.g.][]{Hogg2006,Goto2007,Pawlik2018}, and at $z\sim1$ such 
as the DEEP2 Galaxy Redshift Survey \citep[e.g.][]{Yan2009}. 
In addition, imaging observations have revealed that a significant fraction 
of PSB galaxies present morphological disturbance in their optical images 
signifying an ongoing or past merger, although the merger 
fractions of PSB galaxies in different studies span a wide range 
depending on the merger identification method, PSB sample selection
and imaging data depth and resolution
\citep[e.g.][]{Couch1994,Dressler1994,Zabludoff1996,Caudwell1997,
	Oemler1997,Couch1998,Tran2004,Goto2005,Yang2008,Pracy2009,
	Trouille2013,Pawlik2018,Sazonova2021,Wilkinson2022,Verrico2022}. 

Studies of galaxy structure have found that PSB galaxies at intermediate-to-high redshifts ($0.5<z<2.5$) are more compact than both star-forming and 
quiescent galaxies
\citep[e.g.][]{Newman2012,Yano2016,Almaini2017,Maltby2018,Wu2018,Suess2020,Setton2022}. For instance, \cite{Setton2022} found the PSB galaxies at $z\sim0.7$ 
systematically lie $\sim0.1$dex below the size-mass relation 
of quiescent galaxies. For local galaxies, \cite{Chen2022} have recently 
examined the 
size-relation of the SDSS-based PSB galaxy sample, finding the PSB 
galaxies to be smaller than 
quiescent galaxies of similar mass and star formation rate (SFR), 
especially in the stellar mass range of $\rm10^{9.5}M_\odot\sim10^{10.5}M_\odot$. 
These results are consistent with the picture of 
fast quenching as induced by major mergers. It has been suggested 
that, mergers may drive gas inward and centralized star formation, 
thus leading to a more compact light distribution and resulting in both 
quenching and structural transformation
\citep[e.g.][]{Wellons2015,Zheng2020}.

Although many studies have provided strong evidence for a merger origin 
of PSB galaxies, the fact  that a considerable fraction of PSB galaxies 
show regular morphologies appear to require other mechanisms in 
addition to mergers 
in order to fully explain the PSB galaxy population. As pointed out in 
\citet{Sazonova2021} and \citet{Wilkinson2022}, however, features of 
morphology disturbance fade on timescales of $\sim 200$Myr according 
to hydrodynamic simulations, and so a possible merger origin for all the 
PSB galaxies cannot be simply ruled out based on the merger fractions 
estimated from galaxy images. 

Previous studies mostly used either single-fiber or slitless spectroscopy
which probes only a small region in the galactic center of low-$z$ galaxies
(e.g. the central 1-2 kpc in SDSS galaxies). Integral field spectroscopy
available in recent years has allowed the PSB features to be detected 
in off-center regions of many galaxies
\citep[e.g.][]{Chen2019,Zheng2020,Greene2021,Wu2021,Otter2022,Werle2022}.
Using an early sample of the Mapping Nearby Galaxies at Apache Point 
Observatory
\citep[MaNGA][]{Bundy2015} survey, \cite{Chen2019} identified 360 PSB 
galaxies, of which only 31 galaxies present the PSB feature in their center
(hereafter central PSB galaxies, or cPSB galaxies in short).
In their sample, 37 galaxies show a ring-like PSB region (hereafter rPSB 
galaxies), and the PSB regions of the remaining 292 galaxies have irregular
shapes and locations. Both cPSB and rPSB galaxies in their sample 
presented positive gradients in \dbalmer~(the spectral break at around 
4000\AA) indicative of younger stellar populations in the inner regions.
This result implied that the star formation in PSB galaxies get shut down 
from outside in, opposite to the ``inside-out'' quenching process as 
found for massive galaxies in the general population 
\cite[e.g.][]{Li2015,Wang2018}. The outside-in quenching was also seen 
by \cite{Wu2021} in a small sample of MaNGA-based PSB galaxies, and by 
  \cite{Werle2022} in the PSB galaxies at the centers of eight clusters 
  from $z\sim 0.3$ to $z\sim 0.4$ as observed with the Multi Unit Spectroscopic Explorer (MUSE).

In this work, we make full use of the final data release of the MaNGA survey, 
which provides integral field spectroscopy for $\sim$10,000 nearby galaxies, 
to study both the size-mass relation and the recent star formation history (SFH)
of PSB galaxies in the local Universe. We consider not only the cPSB and 
rPSB types of PSB galaxies as in previous MaNGA-based studies, but also 
those with irregular PSB regions (hereafter iPSB galaxies), which dominate 
the PSB galaxy population. We use three spectral indices to indicate the 
recent SFH of the PSB regions: the equivalent width (EW) of the \ha\ emission line 
(\ewha, indicator of ongoing star formation), the EW of the H$\delta$ 
absorption line (\ewhd, sensitive to young population formed in the past 
0.1-1 Gyr; \citealt{Bruzual2003}), and \dbalmer~(a good proxy for the 
luminosity-weighted stellar age; \citealt{Kauffmann2003}). 
We make comparisons between each 
of the PSB samples and a carefully-selected control sample, closely 
matched in redshift, stellar mass and star formation rate. As we will show, 
the different PSB samples behave differently in both the 
size-mass relation and the SFH diagnostic parameters, indicating 
different evolutionary pathways for different types of PSB galaxies.


This paper is organized as follows. In \autoref{sec:data}, we describe the MaNGA data and the identification of the PSB galaxies. In \autoref{sec:results}, we present the analysis of the size-mass relation and recent SFH.  
We discuss our results in \autoref{sec:discussion} and summarize 
in \autoref{sec:summary}. 
Throughout the paper, we assume a $\Lambda$CDM cosmology with $\Omega_m=0.3$ and $\Omega_{\Lambda}=0.7$, and a Hubble constant of $\rm H_0 = $70 km $\rm s^{-2}$ $\rm Mpc^{-1}$.

\section{Data} \label{sec:data}
\subsection{MaNGA} \label{subsec:manga}

MaNGA is one of the three major projects of the fourth-generation Sloan Digital Sky Survey 
\citep[SDSS-IV;][]{Blanton2017}. During its six-year operation from July 2014 through 
August 2020, MaNGA has successfully obtained high-quality integral field spectroscopy 
(IFS) for 10,010  galaxies with redshift $0.01<z<0.15$ and stellar mass 
$5\times10^8\msun\leq \mstar\leq 3\times10^{11}\msun$ \citep{Bundy2015,SDSS_DR17}. 
MaNGA target galaxies are selected from the SDSS spectroscopic galaxy sample and 
are included in three samples \citep{Yan2016a,Wake2017}: Primary, Secondary and 
Color-Ehanced samples. Targets in the Primary and Secondary samples are each observed 
out to 1.5 or 2.5 times effective radius ($R_e$), and the two samples together have a flat 
distribution of the $K$-corrected $i-$band absolute magnitude ($M_i$). The Color-Enhanced 
sample further selects low-mass red galaxies and high-mass blue galaxies on the plane 
of NUV$-i$ color index versus $M_i$, which are not well sampled by the Primary/Secondary 
samples \citep{Wake2017}. 

The IFS data of MaNGA are obtained using 17 pluggable Integral Field Units (IFUs) with 
a field of view ranging from $12^{\prime\prime}$ to $32^{\prime\prime}$, and each IFU 
is a hexagonal-formatted fiber bundle made from $2^{\prime\prime}$-core-diameter 
fibers with a $0.5^{\prime\prime}$ gap between adjacent fibers, thus producing datacubes 
with an effective spatial resolution of $\sim2.5^{\prime\prime}$  \citep{Drory2015,Law2015}. 
The fibers are fed to two dual-channel BOSS spectrographs \citep{Smee2013} at the 
2.5-meter Sloan Telescope \citep{Gunn2006}, producing spectra with a spectral 
resolution of $R \sim $2000 in the  wavelength range from 3622 to 10354\AA.  A typical 
exposure time of 3 hours ensures a $r$-band continuum S/N of 4$\sim$8 (per-fiber per pixel) 
at 1$\sim$2 $R_e$ \citep{Drory2015}. MaNGA raw data are reduced with the Data 
Reduction Pipeline \citep[DRP;][]{Law2016} to produce sky-subtracted and 
spectrophotometrically-calibrated spectra for scientific studies. Details of the flux 
calibration, MaNGA survey strategy, and data quality tests can be found in \cite{Yan2016a, Yan2016b}. 
All the MaNGA data products are released as part of the final data release of SDSS-IV \citep{SDSS_DR17}.

\begin{figure*}[ht!]
	\centering
	\includegraphics[width=\linewidth]{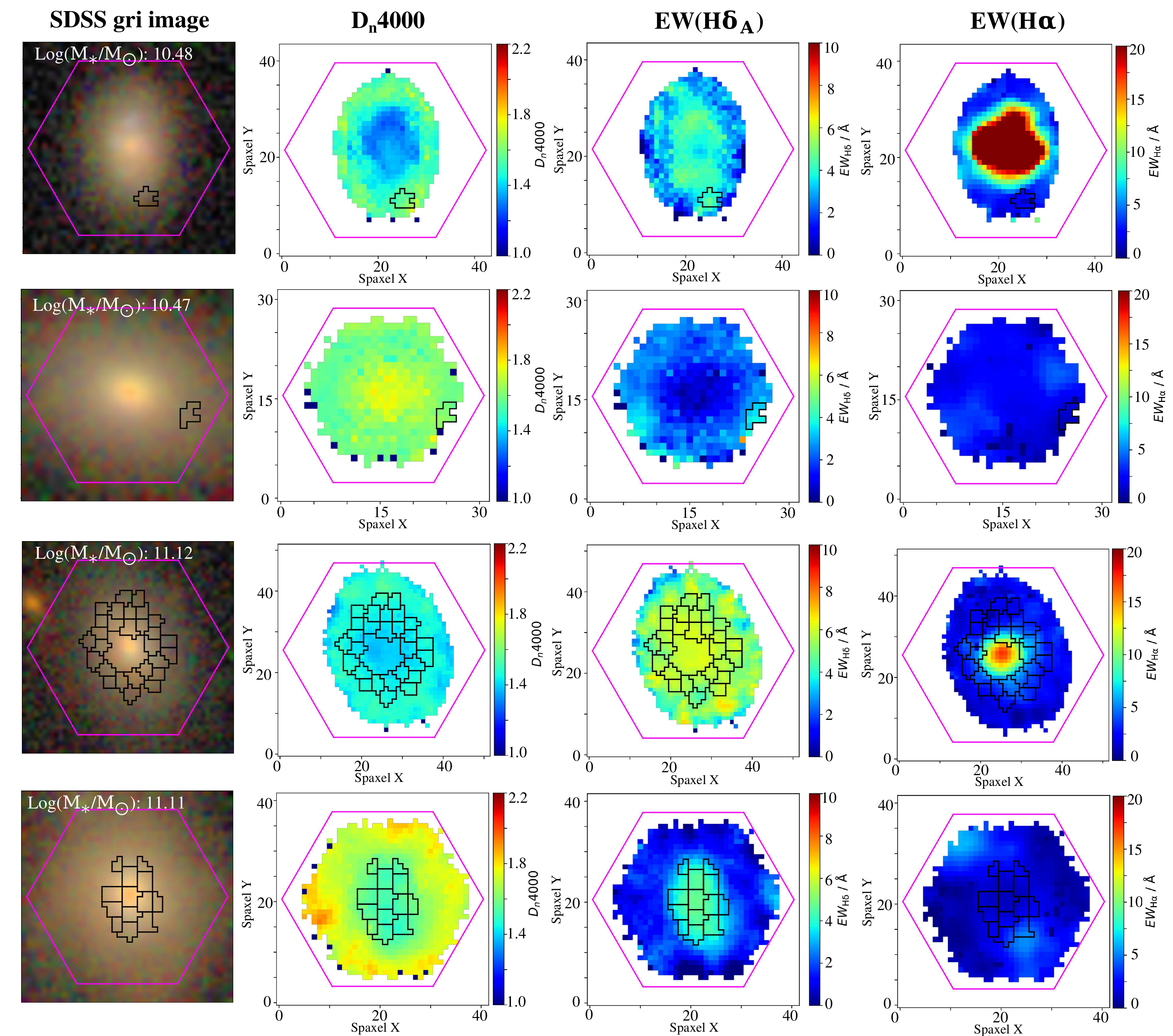}
	\caption{Four examples of PSB galaxy. From left to right, the first column:SDSS image of the galaxy with the magenta hexagonal field of view of MaNGA; the second column: $\ewhd$
		map of this galaxy; the third column: $\ewha$ map of this galaxy; the forth column: $\dbalmer$ map of this galaxy. Black boundaries of
		PSB regions are over-plotted in all panels.}
	\label{fig:psb_example}
\end{figure*}

\subsection{Identification of post-starburst galaxies}\label{subsec:sample}

PSB galaxies are commonly identified by two spectral features jointly: 
strong equivalent width (EW) of high-order Balmer absorption lines produced 
by massive young stars, and weak/no H$\alpha$ emission indicative of weak/no 
ongoing star formation. 
We identify our PSB galaxies by searching for continuous regions in MaNGA
datacubes that present both features. This is done in two successive steps:  we first 
select ``H$\delta$-strong regions'' in each galaxy that have strong H$\delta$ 
absorption, from which we then identify PSB regions by examining both H$\alpha$ 
emission and H$\delta$ absorption lines in the stacked spectrum of each region. 
This procedure is similar to the procedure of identifying Wolf-Rayet galaxies 
in \cite{Liang2020} where the authors firstly search for H{\sc ii} regions in the 
MaNGA datacubes and then  examine the stacked spectra of those regions to 
further identify Wolf-Rayet regions. The reason to use stacked spectra rather than 
original spaxels is two-fold. On the one hand, as shown in \cite{Liang2020}, 
by stacking the spectra of all the original spaxels in a region, we can significantly 
increase the spectral S/N, thus ensuring robust identification for regions with 
relatively weak spectral features. On the other hand, the limited spatial resolution 
of MaNGA ($\sim2.5^{\prime\prime}$, or $\sim1.5$kpc at the median redshift 
$\left<z\right>\sim0.03$ of MaNGA) means that our data cannot resolve individual 
PSB regions, which are smaller than $\sim100$pc if assumed to have sizes 
comparable to H{\sc ii} regions. Therefore, following \cite{Liang2020}, we adopt 
a maximum radius for our PSB regions to be 
$r_p^{\textrm{max}}=\textrm{max}\{1.5^{\prime\prime}, 500\textrm{pc}\}$,
which is comparable to half of the MaNGA resolution, and is a trade-off between 
higher S/Ns from stacking more spectra and stronger dilution to PSB features. 

In practice, for a given galaxy we first perform full spectral fitting to the MaNGA 
spectrum of every spaxel in the datacube using the  spectral fitting code described 
in detail in \cite{Li2020}. The code uses stellar templates constructed by a Principle 
Component Analysis of the single stellar population (SSP) models of \citet{Bruzual2003} and 
assumes a Chabrier initial mass function \citep{Chabrier2003}.
We measure the EW of H$\delta$ absorption line $\ewhd$ from 
the best-fit model spectrum of the stellar component, and the EW of H$\alpha$ 
emission line $\ewha$ from the starlight-subtracted spectrum. We then 
search for H$\delta$-strong regions on the two-dimensional map of $\ewhd$. For this purpose, we have modified {\tt HIIEXPLORER}, a 
pipeline developed by \cite{Sanchez2012} for the identification of {\sc Hii} regions
 on maps of H$\alpha$ surface density. The modified code starts by picking 
 up the spaxel with the highest $\ewhd$ ($>3$\AA) over the map as the center 
 of a potential ``H$\delta$-strong region''.  Adjacent spaxels in the vicinity of 
 the central spaxel are appended to the region if their distances from 
 the central spaxel are less than $r_p^{\textrm{max}}$. The region is removed 
 from the map and the above procedure is repeatedly applied to the remaining 
 map until every spaxel is assigned to a region, or rejected. 

For each of the H$\delta$-strong regions, we stack the spectra of its spaxels
in the same way as in \cite{Liang2020}. In short, the spaxel spectra are corrected 
to the rest frame using both the galaxy redshift and the stellar velocity of each 
spaxel relative to the galaxy, and are weighted by the inverse variance of flux errors
provided by the MaNGA DRP \citep{Law2016} to produce a stacked spectrum 
of the region. Flux errors of the stacked spectrum are firstly calculated by the 
standard formula in weighted mean statistics, and then corrected for the effect 
of covariance following the formula in \citet[][see their Figure 16]{Law2016}. 
We ignore regions with the stacked spectrum of S/N$<10$ in what follows.
The code of \cite{Li2020} is used again to fit each of the stacked spectra 
(with S/N$>10$), from which we measure all emission line parameters and 
stellar indices that may be used in our work. We identify PSB regions from the
H$\delta$-strong regions, following the same selection criteria as 
in \citet[][see their \S2.3]{Chen2019}: 
(i) EW(H$\delta_{A}$)$>$ 3\AA,
(ii) EW(H$\alpha$) $<$ 10\AA, and 
(iii) $\log_{10}$[EW(H$\alpha$)]$<$ 0.23 $\times$ EW(H$\delta_{A}$) $- 0.46$.
We note that we have adopted selection criteria from other studies 
\citep[e.g.][]{Goto2008, Chen2022} and repeated the whole analysis, 
finding our scientific conclusions to be robust to the different selection criteria.

A galaxy is identified to be a PSB galaxy if it contains one or more PSB regions. 
In total, we have identified 3204 PSB regions, distributed in 489 PSB galaxies. 
Following \cite{Chen2019}, we visually examine the maps of EW(H$\delta_A$)
and EW(H$\alpha$) and classify our PSB galaxies into three types according 
to the location of their PSB regions: (i) cPSB with the PSB regions located at the 
galactic center, (ii) rPSB with a ring-like distribution of multiple PSB regions,  
and (iii) iPSB with the PSB regions distributed in an irregular manner. 
The numbers of the three types of galaxies are 310 (iPSB), 85 (rPSB) and 94 (cPSB).
Thus the majority of PSB galaxies are iPSB, which have been rarely studied, however.
In \autoref{fig:psb_example} we show four example PSB galaxies in our sample: 
two iPSB galaxies in the top two rows (one star-forming galaxy and one quiescent galaxy according to their global star formation rate), 
followed by a rPSB galaxy and a cPSB galaxy. The four panels (from left to right) 
show the SDSS $gri$ composite image and maps of 
\dbalmer (the narrow version of the spectral break at around 4000\AA, 
as originally defined in \citealt{Balogh1999}), \ewhd\ and \ewha. 
The latter three parameters are sensitive to young stellar populations of 
different ages, thus can be used as good diagnostics for the recent 
star formation history (SFH) of given region \citep[e.g.][]{Li2015,Wang2018}. 
The different types of PSB galaxies differ from 
	each other in these diagnostic parameters, indicating diverse SFHs 
in PSB galaxies. We will study the recent SFHs of our PSB galaxies in 
the next section.

\section{Results} \label{sec:results}

\subsection{Global properties of PSB galaxies}
\label{sec:globalproperties}

\begin{figure*}[ht!]
	\includegraphics[width=\linewidth]{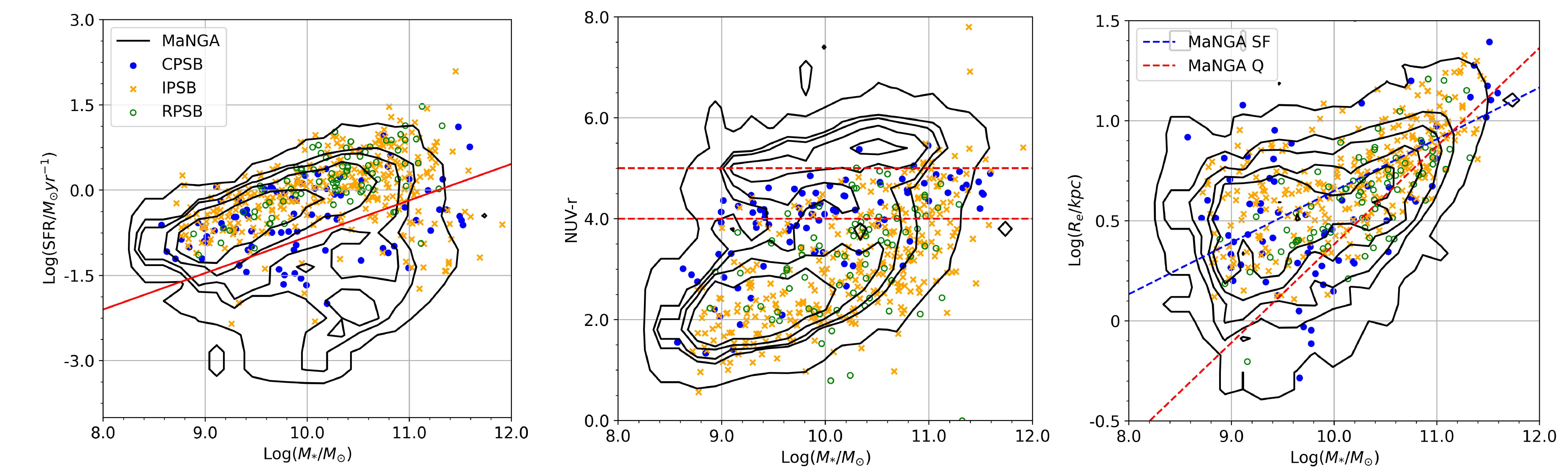}
    \caption{$Left$ $panel$: cPSB(blue dots), rPSB(green circles) and iPSB(yellow crossed) are plotted on the stellar mass versus SFR diagram. The black contour represents the MaNGA parent sample. The sample selection effects has been corrected with $1/V_{max}$ weighting method. More details can be found in \cite{Wake2017}. The red solid line is the empirical divided line of SFG and QG. $Middle$ $panel$: Same as left panel, but showing stellar mass versus NUV-r color. The two horizontal dashed lines indicate NUV-r = 4 and 5. $Right$ $panel$: Same as left panel, but showing stellar mass versus $r$-band effective radius.}
   \label{fig:sfr_m_distribution}
\end{figure*}

\autoref{fig:sfr_m_distribution} examines the global properties of our PSB 
galaxy sample by displaying all the PSB galaxies in three diagrams:
the SFR-mass diagram (left), the color-mass diagram (middle) and the size-mass 
relation (right), with different colors/symbols for the three PSB types.
For comparison, the distribution of the parent MaNGA sample is plotted as 
black contours, for which we have corrected the effect of sample incompleteness
using the $1/V_{\textrm{max}}$-weighting scheme following \citet{Wake2017}. 
Stellar masses ($M_\ast$) and star formation rates (SFR) of our galaxies are taken 
from the GSWLC catalog\footnote{\url{https://salims.pages.iu.edu/gswlc/}}
\citep{Salim2016}, and effective radii ($R_e$) are provided in the NSA,
measured through two-dimensional Sersic profile fits to SDSS $r$-band images
\citep{Blanton2011}. 
The NUV$-r$ color indices, from NSA catalog, are based on 
the NUV image from GALEX and the $r$-band image from SDSS. 
The red solid line in the SFR-mass diagram is  the empirical divider 
from \cite{Woo2013}, $\log_{10}$SFR($\rm M_{\odot}$ ${\rm yr}^{-1}$) = 0.64 
$\times$ log $\rm M_{*}(M_{\odot}) - 7.22$,
which separates star-forming (SF) galaxies from quiescent (Q) galaxies. 
The two horizontal dashed lines in the color-mass diagram represent 
the commonly-used cuts of NUV$-r$=4 and NUV$-r$=5
which divide the galaxies into red, green and blue populations.
The dashed lines in blue and red in the right panel are linear fits 
to the $R_e$-$\rm\log_{10}M_\ast$ relation of star-forming galaxies 
and quiescent galaxies, as selected from the parent MaNGA sample
on the SFR-mass diagram. 

Overall, to our surprise, the PSB galaxies in our sample are rarely red or 
quenched.  Rather, they are mostly blue or green with NUV$-r<5$ 
in the color-mass diagram, and are mostly located in the star-forming 
sequence or the transition region above the quiescent sequence in 
the SFR-mass diagram. This is in contrast to the traditional view
of PSB galaxies which have long been assumed to be quenched 
after having experienced a recent starburst. When divided into the 
three types, the PSB galaxies present different distributions in these 
diagrams. First, the majority of the cPSB galaxies are found in the 
transition region between red/blue or SF/Q populations, thus 
consistent with the traditional expectation. Quantitatively, out of the 
94 cPSB galaxies in our sample, 48 (51.1\%) fall in the green valley 
with $4<$NUV$-r<5$, and  27 (28.7\%) are located just below the 
star-forming sequence with only a few scattering to the sequence
of quiescent galaxies. This result is similar to previous
findings from SDSS-based studies where the PSB galaxies 
were identified based on spectroscopy of the central 1-2 kpc
\citep[e.g.][]{Wong2012,Chen2022}, and it is broadly consistent 
with the standard conjecture that PSB galaxies are in the transition 
phase between star-forming and quenched galaxies.  On the other 
hand, however, the spatially resolved spectroscopy from MaNGA 
reveals that more than two third of the cPSB galaxies are still globally 
star-forming and blue, although the star formation in their central 
region has recently shutdown. Obviously, even when limited to galactic 
centers, a significant fraction of PSB galaxies are not fully quenched. 

Next, different from cPSB galaxies, the rPSB galaxies are mostly located 
in the SF sequence and the blue cloud, with only 3.5\% (3/85) 
falling below the SF sequence and 25.9\% (22/85) having NUV$-r>4$.
In an earlier MaNGA-based study by \citet{Chen2019}, cPSB galaxies 
are found in the transition region in the plane of $\dbalmer$   versus $\rm\log_{10}M_\ast$, and rPSB galaxies fall slightly below 
the cPSB population but above the sequence of the young 
galaxy population (see the bottom-left panel of their Figure 5). 
Our result is not inconsistent with theirs, considering that the 
$\dbalmer$ examined in their work, the SFR we used and NUV$-r$ 
examined here are \textcolor{black}{sensitive to stellar populations formed at different times.} 
We will examine the $\dbalmer$ of our galaxies in \autoref{sec:sfh}.
Finally, different from both cPSB and rPSB galaxies which are 
limited to narrow ranges of SFR and NUV$-r$, the iPSB galaxies 
cover a wide range in both parameters, though with a relatively 
large fraction falling in the star-forming sequence ($269/310=86.8\%$)
and the blue cloud with NUV$-r<4$ ($235/310=75.8\%$). 

\textcolor{black}{As estimated by applying 
	the SED fitting technique to multiband photometry, the SFRs from 
	GSWLC are averaged over the past 30-100 Myr and so they may 
	be higher than the current SFRs for PSB galaxies for which the SFRs 
	are expected to have rapidly declined. We have attempted to estimate the 
	global SFRs for our galaxies using the total H$\alpha$ luminosity within 
	1.5$Re$ from the MaNGA data and compared the measurements 
	with those from GSWLC. This comparison has to be limited 
	to a fraction of the MaNGA galaxies that have substantially high 
	SNR in most of the spaxels. Basically, we found consistent SFR 
	measurements with the GSWLC with no systematics over the full 
	SFR range, though with a median offset of 0.11 dex 
	in $\log_{10}{\rm SFR}$ due to the limited spatial coverage of the
	MaNGA datacubes. More importantly, the PSB galaxies show similiar 
	results to the parent MaNGA sample in this comparison, demonstrating 
	that the different timescales for SFR between SEDs and 
	H$\alpha$ luminosities make little difference in our analyses of the 
	PSB galaxies. This can be understood from the fact that, PSB regions
	happen only in parts of the host galaxies, thus contributing only to a fraction 
	of the global SFR.
}

In the size-mass diagram as shown in the rightmost panel of 
\autoref{fig:sfr_m_distribution}, the parent sample shows a \textcolor{black}{broad}
distribution, and at fixed mass the populations with larger 
and smaller sizes roughly correspond to the star-forming 
and quiescent galaxy populations. This result echoes early studies 
of the size-mass relation of local galaxies \citep[e.g.][]{Shen2003},
where galaxies with early-type and late-type morphologies 
showed distinct size-mass relations. As one can see,
the PSB galaxies in our sample cover the full area
of the parent MaNGA sample except the regime of smallest $R_e$
at fixed stellar mass, which are dominated by quiescent galaxies 
with smaller-than-average sizes. This is consistent with the 
left two diagrams where the PSB sample as a whole lacks the 
most quiescent or the reddest galaxies. When divided into different types, 
cPSB galaxies with intermediate masses ($9.5\lesssim$\lgmstar$\lesssim10.5$)
appear to be smaller than other types of PSB galaxies which 
follow more closely the parent sample. At lower masses, 
all types of PSB galaxies show similar distributions to the parent sample.
We will further study the size-mass relation 
for the different types of PSB galaxies in the following subsection.

\begin{figure*}[htbp]
	\centering
	\includegraphics[width=\linewidth]{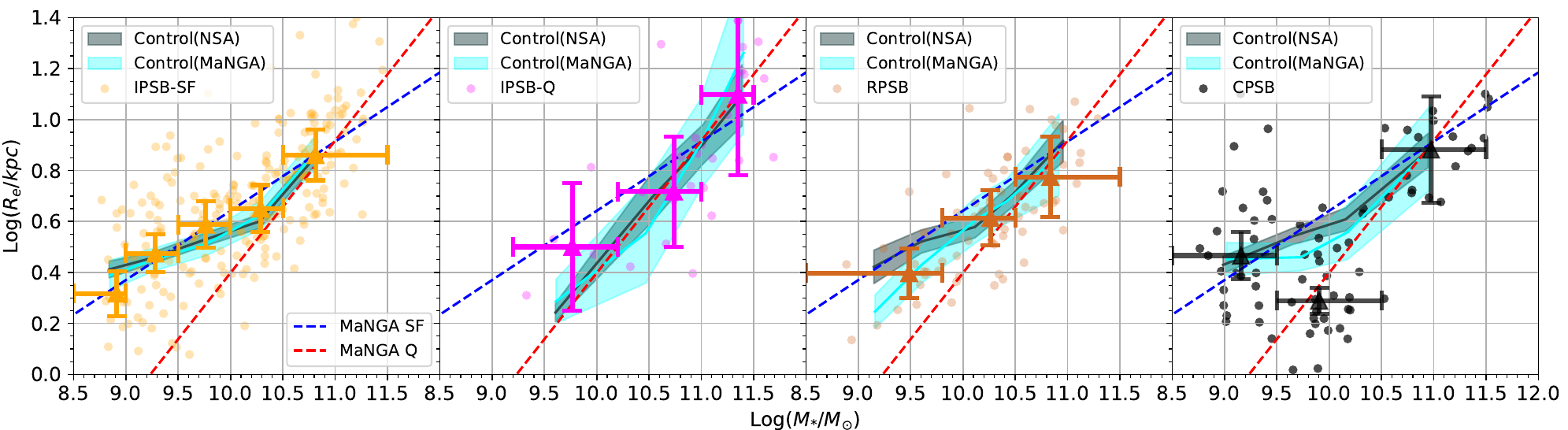}
        \includegraphics[width=\linewidth]{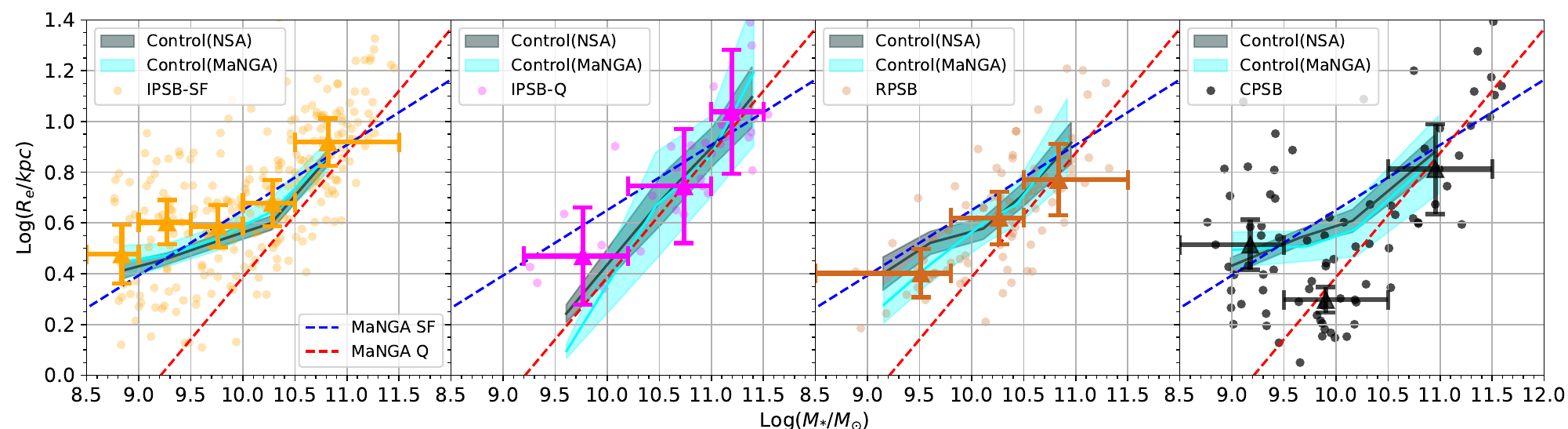}
        \includegraphics[width=\linewidth]{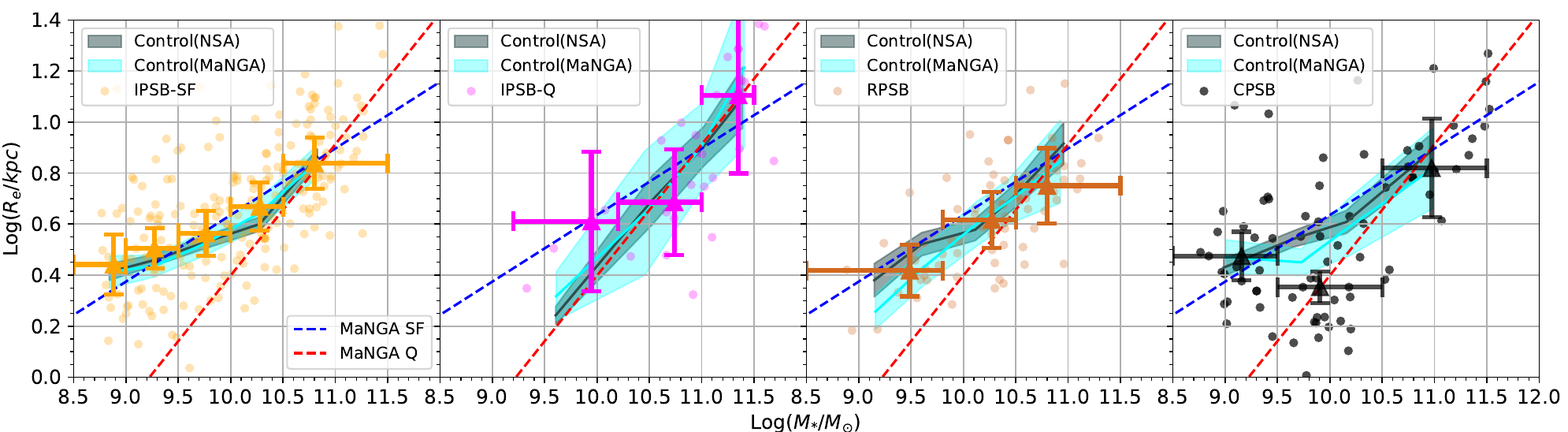}
    \caption{Size-mass relation for PSB galaxies and control samples. Panels from left to right are four PSB subsamples:iPSB-Q(purple),  iPSB-SF(orange), rPSB(brown) and cPSB(black). Panels from top to bottom show $R_e$ measurements based on SDSS $g$-band, $r-$band and $i-$band image. The dot symbols show the distribution of individual PSB galaxies. The median values in different stellar mass bins are plotted. Error bar indicates the Poisson error. The median values of control SFR samples are also presented in each panel. The shadow regions denote the Poisson errors. The dashed line is the best fit for the size–mass relation of MaNGA galaxies in formula of log $R_e$(kpc) = k $\times$ log $\rm M_{*}$($\rm M_{\odot}$) + b. }
    \label{fig:size_mass}
\end{figure*}

\begin{figure*}[htbp]
	\includegraphics[width=\linewidth]{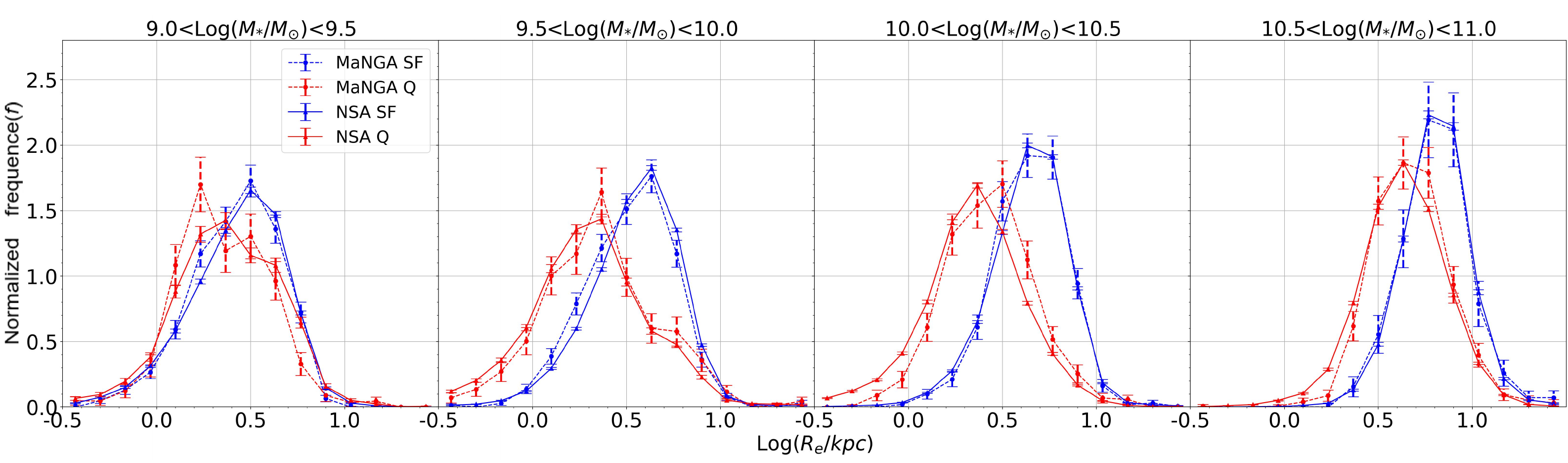}
	\caption{The distributions of log $R_e$(kpc) in different stellar mass bins for MaNGA(dashed line) and NSA volume-limited sample(solid line). Error bars are Poisson counting errors. Sample selection effects with the $1/V_{max}$ weighting method have been considered during our calculation for the MaNGA sample.}
	\label{fig:compare}
\end{figure*}

\subsection{The size-mass relation}
\label{sec:size-mass}


In \autoref{fig:size_mass} we show the size-mass relation again, but for the 
different types of PSB galaxies separately (panels from left to right) and 
for sizes measured in three different photometric bands (panels from top to bottom). 
As we have seen from \autoref{fig:sfr_m_distribution}, the 
size-mass relation strongly depends on the global star formation rate. 
Considering the wide coverage of the iPSB galaxies in the SFR-mass diagram, 
we divide them into two subsets according to their location in the diagram. 
This division results in 269 ``iPSB-SF'' galaxies located above the dividing 
line (solid red line in the left panel of \autoref{fig:sfr_m_distribution}) and 
41 ``iPSB-Q'' galaxies located below that line. For cPSB and rPSB samples 
we don't attempt to further select subsamples considering both the limited sample 
sizes and the fact that they occupy relatively narrow ranges in SFR and color 
at fixed stellar mass. To take into account the 
dependence on mass and SFR and to make fair comparisons with normal galaxies, 
for each of the PSB samples we follow \citet{Chen2022} to select a ``control'' sample
of galaxies from the MaNGA sample that is closely matched with the PSB galaxies 
in redshift, $M_\ast$ and SFR. In practice, we require the control sample to contain 
at most three nearest-matched galaxies for each PSB galaxy, with matching 
tolerances of \textcolor{black}{
$|\Delta z|<0.005$, $|\Delta \rm\log_{10}M_\ast|<0.05$ and $|\Delta\rm\log_{10}SFR|<0.05$.}
This gives rise to sufficient matches for 80\% of the PSB galaxies in our sample. 
For the remaining 20\% galaxies, we relax the tolerances to  be \textcolor{black}{
$|\Delta z|<0.01$, $|\Delta \rm\log_{10}M_\ast|<0.2$ and 
$|\Delta\rm\log_{10}SFR|<0.2$} in order to have at least one match for each PSB galaxy.

In \autoref{fig:size_mass}, panels from left to right compare the size-mass relation
for PSB galaxies of type iPSB-SF, iPSB-Q, rPSB and cPSB with that of the
corresponding control samples. Panels from top to bottom show the results 
for $R_e$ measured in SDSS $g$, $r$ and $i$ band. 
For PSB samples, we show both individual galaxies
in each sample (as the colored dots) and the median and $1\sigma$ scatter 
(as the big crosses), while for the control samples we only show the median and 
scatter for clarity (as the cyan band). The dashed lines in blue and red present the 
result for the star-forming and quiescent galaxies in the parent sample, and are 
repeated in every panel for reference. We find that, when divided by the PSB type, 
the PSB galaxies in our sample show different size-mass relations.
The iPSB-SF and iPSB-Q types respectively follow the relations of the star-forming 
population (the blue dashed line in each panel) and the quiescent 
population (the red dashed line in each panel), while the rPSB type appears to 
fall in between. The cPSB galaxies follow the star-forming population at both high
and low masses, but have relatively smaller sizes at intermediate masses
($9.5\lesssim$\lgmstar$\lesssim 10.5$) which are comparable to the sizes
of the quiescent population. When compared to control galaxies selected
from the parent MaNGA sample, the iPSB-SF, iPSB-Q and rPSB samples closely 
follow the relations of their control samples, while the cPSB galaxies with 
intermediate masses have a median size that is systematically smaller than 
that of the control sample. This result of the cPSB sample is well consistent 
with \citet[][see their Figure 4]{Chen2022}\footnote{We notice that the stellar masses and SFRs 
	used in \citet{Chen2022} were from the MPA-JHU SDSS catalog. 
	We have repeated our analysis using the same catalog and find similar results.}, where 
the SDSS-based PSB galaxies (thus similar to our cPSB galaxies in definition)
are found to be smaller than control galaxies, also at 
$9.5<$\lgmstar$<10.5$. 

Although the MaNGA target galaxies are selected only by redshift and 
$i$-band luminosity, the limited IFU sizes and the requirement for the galaxies
to be covered by the IFUs out to a fixed radius (1.5$R_e$ or 2.5$R_e$)
may have introduced size-related biases in the MaNGA sample. 
In \autoref{fig:compare} we check this out by comparing the distribution 
of $R_e$ of the MaNGA galaxies in narrow bins of $\rm M_\ast$ with the 
distribution of galaxies in a volume-limited sample, which is selected 
from NSA to have $0.01<z<0.08$ and $\rm M_\ast>10^{9} M_{\odot}$. 
For MaNGA galaxies, the distributions are corrected for sample incompleteness
with the $1/V_{\textrm{max}}$-weighting method as mentioned above.
At all masses, we see no significant differences in the size distributions,
indicating that the MaNGA sample is unbiased in terms of galaxy size,
and so our results on the size-mass relations presented above should be
statistically robust and reliable. In addition, we have constructed another 
control sample from the NSA using the same tolerances, and we show 
the size-mass relations of this sample in \autoref{fig:size_mass} as 
grey-shaded region. As can be seen, the two control samples agree 
well with each other, indicating again that the MaNGA sample provides 
unbiased results. Finally, size measurements from different 
bands lead to very similar size-mass relations for any given sample 
as can be seen in \autoref{fig:size_mass}, demonstrating that our results 
are robust to the adopted photometric band.

\begin{figure*}[htbp]
	\centering 
	\includegraphics[width=\textwidth]{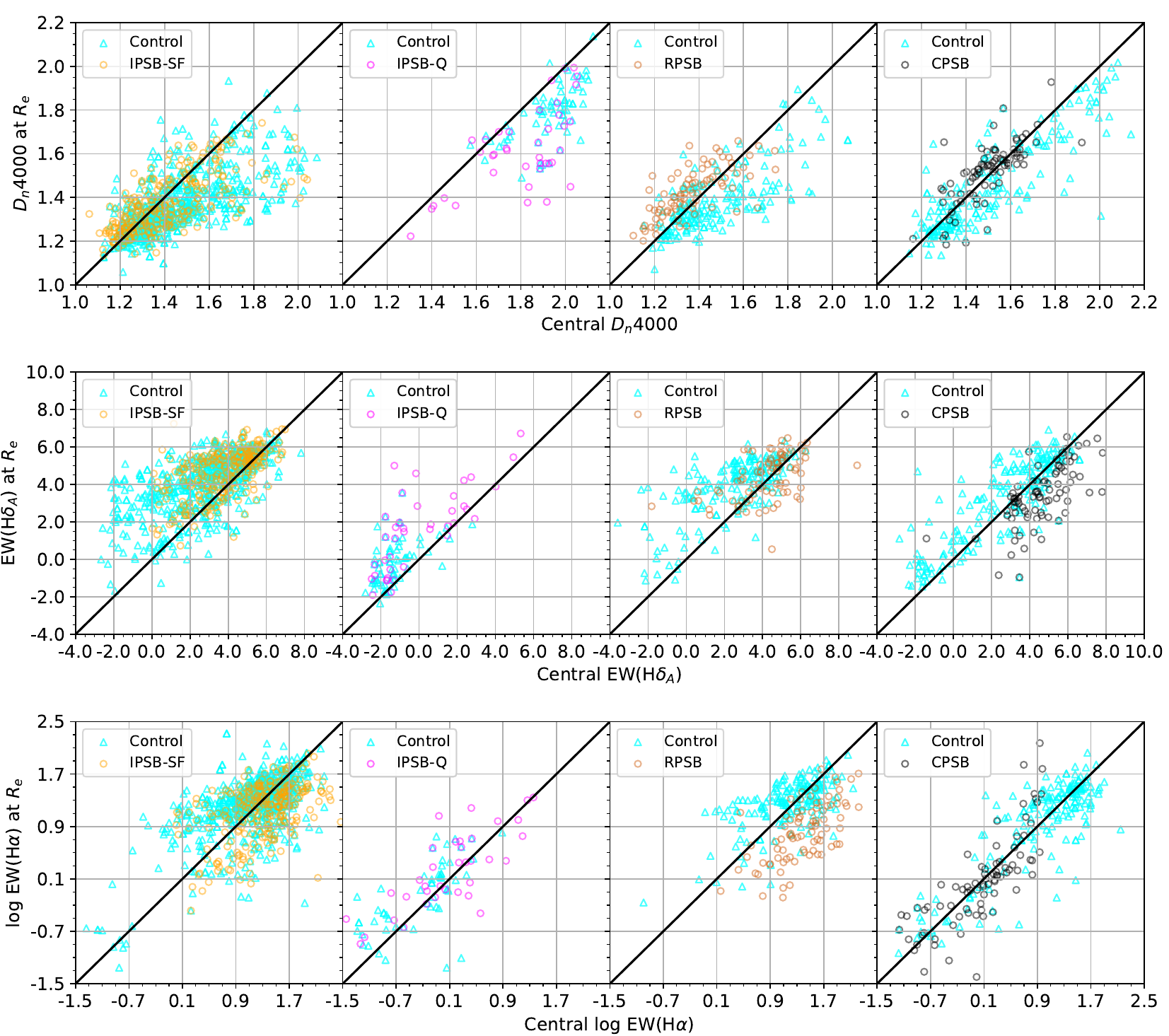}
	\caption{$Top$ $row$: galaxies are plotted on the plane of central $\dbalmer$ versus the $\dbalmer$ at $R_e$. Panels from left to right are four PSB subsamples and corresponding control samples. The colors of symbols are similar to Figure~\ref{fig:size_mass}. The black solid line in each panel represents the 1:1 relation. $Middle$ $row$: same as the top row, but showing the $\ewhd$ result. $Bottom$ $row$: same as the top row, but showing the log$_{10}$EW(H$\alpha$) result.}
	\label{fig:cen_re}
\end{figure*}

\begin{figure*}
	\centering
	\includegraphics[width=\textwidth]{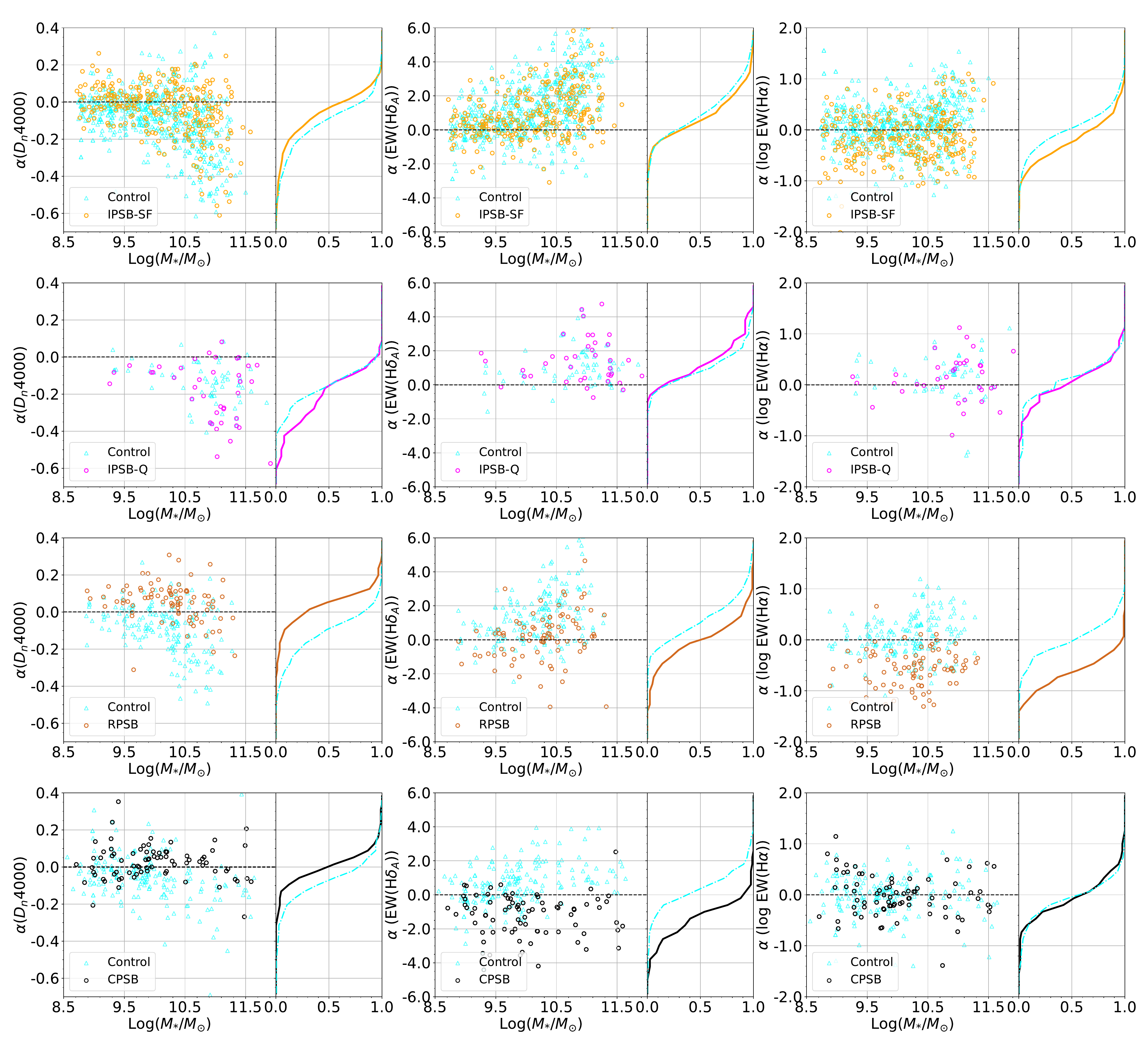}
	\caption{Distributions in the plane of slope index versus stellar mass. Panels from left to right are different diagnostics of recent SFH: $\dbalmer$, $\ewhd$, and $\ewha$. Panels from top to bottom are different PSB subsamples and corresponding control SFR samples. The colors of symbols are similar to Figure~\ref{fig:cen_re}. On the right side of each panel, we also show the cumulative distribution of the slope index.}
	\label{fig:alpha}
\end{figure*}

\subsection{Recent star formation histories}
\label{sec:sfh}


Following \cite{Li2015} and \citet{Wang2018}, we use \dbalmer, \ewhd\ and
\ewha\ to indicate the recent star formation history (SFH) of the 
PSB galaxies and the control galaxies. It is well known that 
\dbalmer\ and \ewhd\ are sensitive to young stellar populations 
formed in the past 1-2 Gyr and 0.1-1 Gyr, respectively, and that \ewha\ 
indicates the ongoing star formation. Therefore, the three parameters combine 
to provide a powerful diagnostic of the recent SFH. By comparing the three 
parameters in the center and outskirt of MaNGA galaxies, \citet{Li2015} and 
\citet{Wang2018} found massive galaxies with $\rm M_\ast\ga 10^{10}M_\odot$ 
to present significant gradients in these parameters, indicative of an
``inside-out'' process for the star formation cessation, while less massive 
galaxies show weak or no gradients. In \autoref{fig:cen_re}, we compare the 
three parameters as measured at $R_e$ and in the central region for the four 
types of PSB galaxies. Panels from left to right correspond to samples of 
iPSB-SF, iPSB-Q, rPSB and cPSB, and panels from top to bottom are the 
results for \dbalmer, \ewhd\ and \ewha. The corresponding 
control galaxies are also shown in each panel for comparison. 

Overall, the different types of PSB galaxies behave differently. 
Most of the iPSB-SF galaxies present no or weak gradients in all 
the three parameters, with the central and outer regions showing 
similar values in all three parameters, which are consistent with 
the global star-forming status of the galaxies. Differently, 
most of the iPSB-Q galaxies present significant gradients in 
\dbalmer\ and \ewhd, with larger \dbalmer\ and smaller \ewhd\ 
(thus older stellar populations) in their center than in the outskirt, 
and their \ewha\ are comparably low ($<10$\AA) in both central 
and outer regions. This result is consistent with the global quiescence 
of the iPSB-Q galaxies, and also consistent with the inside-out quenching 
picture as found previously for the general population of massive galaxies, 
in which the star formation cessation occurs first in the galactic 
center before extending to larger radii. We note that a small fraction 
of the iPSB-SF galaxies with relatively large \dbalmer\ and relatively 
small \ewhd\ appear to deviate from the 1:1 relations, with the central 
region showing larger \dbalmer\ and smaller \ewhd\ than the outskirt. 
Although still forming stars globally, these galaxies are likely to have 
just started the star formation cessation process, expected to further 
deviate from the 1:1 relation and follow up the iPSB-Q galaxies in the future. 
Eventually, both iPSB-SF and iPSB-Q galaxies will be fully quiescent, 
moving back to the 1:1 relation but with largest \dbalmer\ and smallest 
\ewhd\ and \ewha. This result implies that the two types of iPSB galaxies 
follow a similar evolutionary pathway (i.e. the inside-out quenching process), 
but they are currently at different stages. 

The rPSB and cPSB types appear to follow an evolutionary pathway different
from the iPSB galaxies. In most cases, rPSB galaxies have small \dbalmer\ 
($\lesssim 1.6$) and large \ewhd\ ($\ga 3$\AA) in both central and 
outer regions, as well as large \ewha\ ($\ga 10$\AA) in the central 
region. This is consistent with the fact that this type of PSB galaxies 
mostly fall in the star-forming sequence in the SFR-mass diagram 
(see \autoref{fig:sfr_m_distribution}). It is interesting that rPSB 
galaxies show smaller \dbalmer\ and larger \ewha\ in the center 
than in the outskirt, an effect that is opposite to any other type of galaxies.
This indicates older populations and weaker star formation in the 
outskirt, and thus implies an ``outside-in'' quenching mode, if 
we assume these galaxies are starting to cease star formation 
as indicated by the presence of PSB regions in their outskirt. 
cPSB galaxies have small \dbalmer\ ($\lesssim 1.6$) and small 
\ewha\ ($\lesssim 10$\AA) in both central and outer regions, 
implying that these galaxies have ceased star formation 
in the recent past (within 1-2 Gyr). Their central regions have 
larger \ewhd\ ($\ga 3$\AA) than the outer regions, implying that 
the star formation cessation occurs later in the central region 
than in the outskirt. These behaviors are also consistent with the 
``outside-in'' quenching picture. Given the global star-forming 
status of the rPSB galaxies and the quiescent status of the cPSB
galaxies, it is natural to conclude that the two types of PSB galaxies 
follow the outside-in mode of quenching, but are currently at 
different stages. 

The distribution of the control galaxies varies from panel to panel, 
which can be understood by the dependence on stellar mass 
as found in \citet{Wang2018}. 
When compared to the control samples, both iPSB-SF and iPSB-Q
galaxies are similar to (though not exactly the same as) the 
corresponding control galaxies, while rPSB and cPSB galaxies 
are quite different from their control samples.  
This is more clearly seen in \autoref{fig:alpha}, where we plot 
the gradient of the three diagnostic parameters as a function of 
\lgmstar\ for both the PSB samples and the control samples. 
The gradient of a parameter is defined as the difference of its
value at $R_e$ from that in the center. On the right side of each panel, we also show the cumulative distributions 
of the diagnostic parameters. As can be seen, for a given parameter
all the control samples show similar trends between the parameter 
gradient and the stellar mass, with a weak dependence on the 
type of the PSB galaxies. Consistent with \citet{Wang2018},
control galaxies at \mstar$\ga 10^{10}M_\odot$ present significant 
gradients in \dbalmer\ and \ewhd, and no/weak gradients at lower masses. 
Compared to the control samples, both iPSB-SF and iPSB-Q 
galaxies show high similarities. In contrast, both rPSB and cPSB 
galaxies show different gradients in all parameters except 
\ewha\ in case of the cPSB sample. These results suggest 
again that the PSB galaxies in our sample can be divided into 
two broad categories according to their evolutionary pathway:
(i) iPSB-SF and iPSB-Q to follow the inside-out quenching mode
as found for the majority of the general population galaxies, and 
(ii) rPSB and cPSB to follow the outside-in quenching mode.

\color{black}
\subsection{The fraction of PSB regions}
\label{subsec:fpsb}

\begin{figure}
	\centering
	\includegraphics[width=\linewidth]{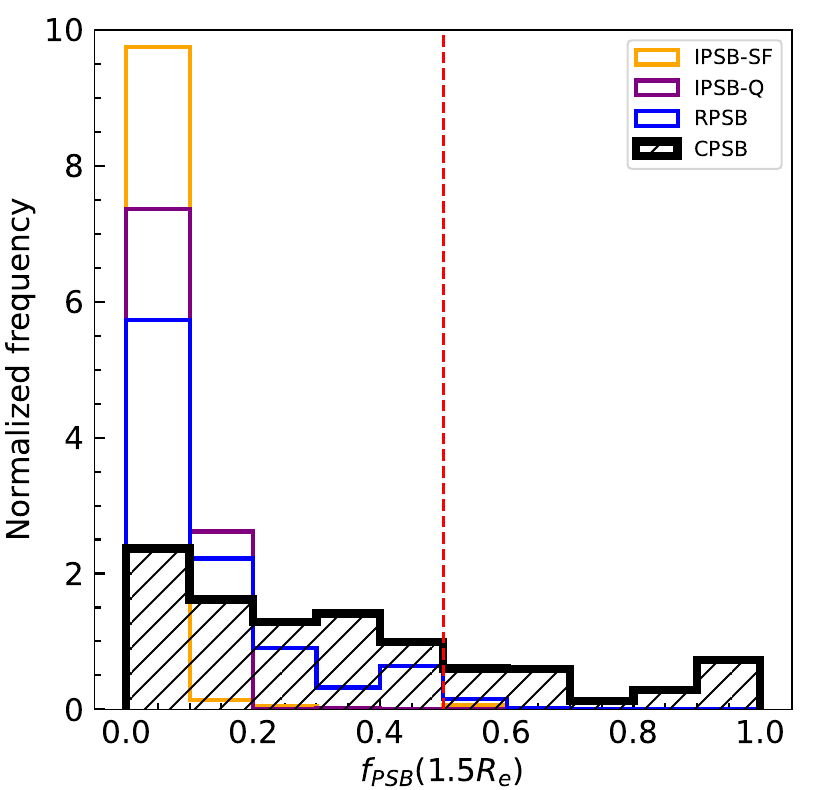}
	\caption{\textcolor{black}{Normalized distributions of $f_{\rm PSB}(1.5R_e)$ for the four types of PSB samples. The 1/$V_{max}$-weighting scheme is performed to correct for sample incompleteness. 
			The red vertical dashed line indicates $f_{\rm PSB}(1.5R_e) = 0.5$.}}
	\label{fig:fig8}
\end{figure}

In this subsection, we further consider the potential dependence of the 
size-mass relation and recent SFH on the fraction of PSB regions in our galaxies. 
The fraction of PSB regions, $f_{\rm PSB}(R)$, is defined by the mass-weighted fraction 
of spaxels that belong to the PSB regions within a given galactic radius $R$,
\begin{equation}
    f_{\rm PSB}(R) = \frac{\sum_{r<R} m_{*,i} \times f_i}{\sum_{r<R} m_{*,i}}
\end{equation}
Here $m_{*,i}$ is the stellar mass of the $i$th spaxel, $f_i=1$ or $0$ depending 
on whether the $i$th spaxel belongs to any PSB region, and $R = 1.5 R_e$. 
\autoref{fig:fig8} shows the histograms of $f_{\rm PSB}(1.5R_e)$ for the different PSB samples. 
We see that both iPSB and rPSB galaxies exclusively have relatively small fractions 
of PSB regions with $f_{\rm PSB}(1.5R_e)<0.5$, while the cPSB galaxies span a 
full range of $f_{\rm PSB}(1.5R_e)$. We find 23.3\% of the cPSB sample to 
have $f_{\rm PSB}(1.5R_e)>0.5$, after the $1/V_{\rm max}$-weighting scheme is 
applied to correct for sample incompleteness.

\begin{figure*}
    \centering
    \includegraphics[width=\textwidth]{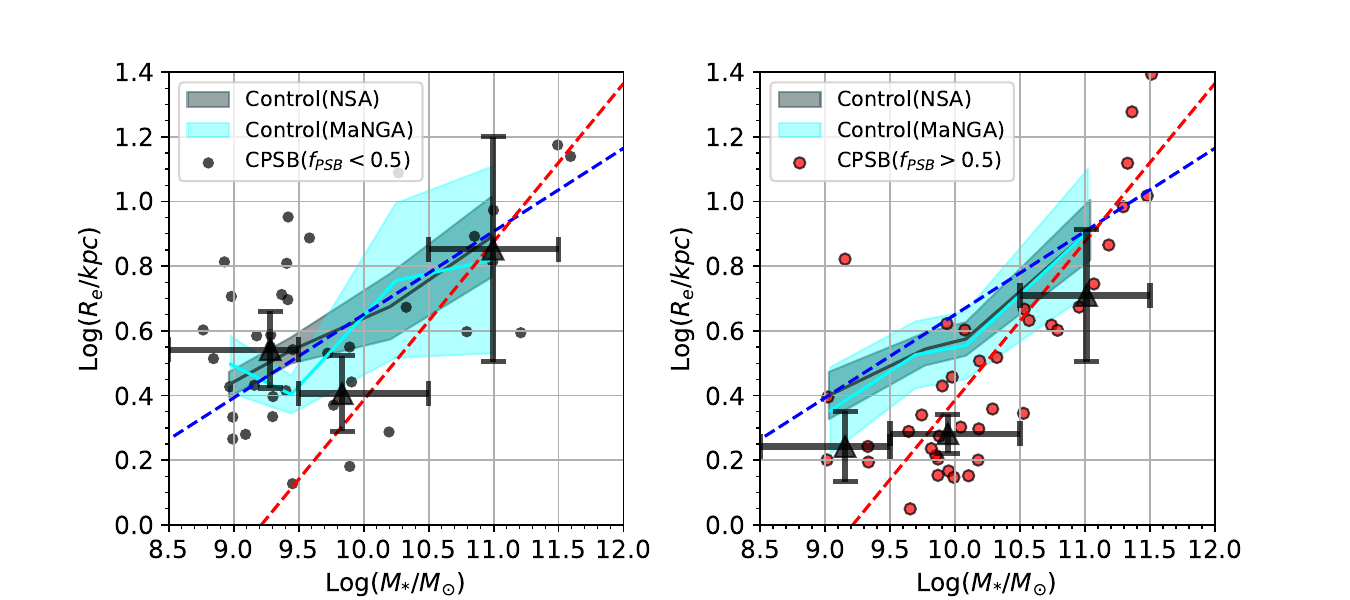}
    \caption{\textcolor{black}{Size-mass relation for cPSB galaxies with $f_{\rm PSB}(1.5R_e)<0.5$ (left panel) and those with $f_{\rm PSB}(1.5R_e)>0.5$ (right panel), as well as the corresponding control samples.}}
    \label{fig:fig9}
\end{figure*}

\begin{figure*}
    \centering
    \includegraphics[width=\textwidth]{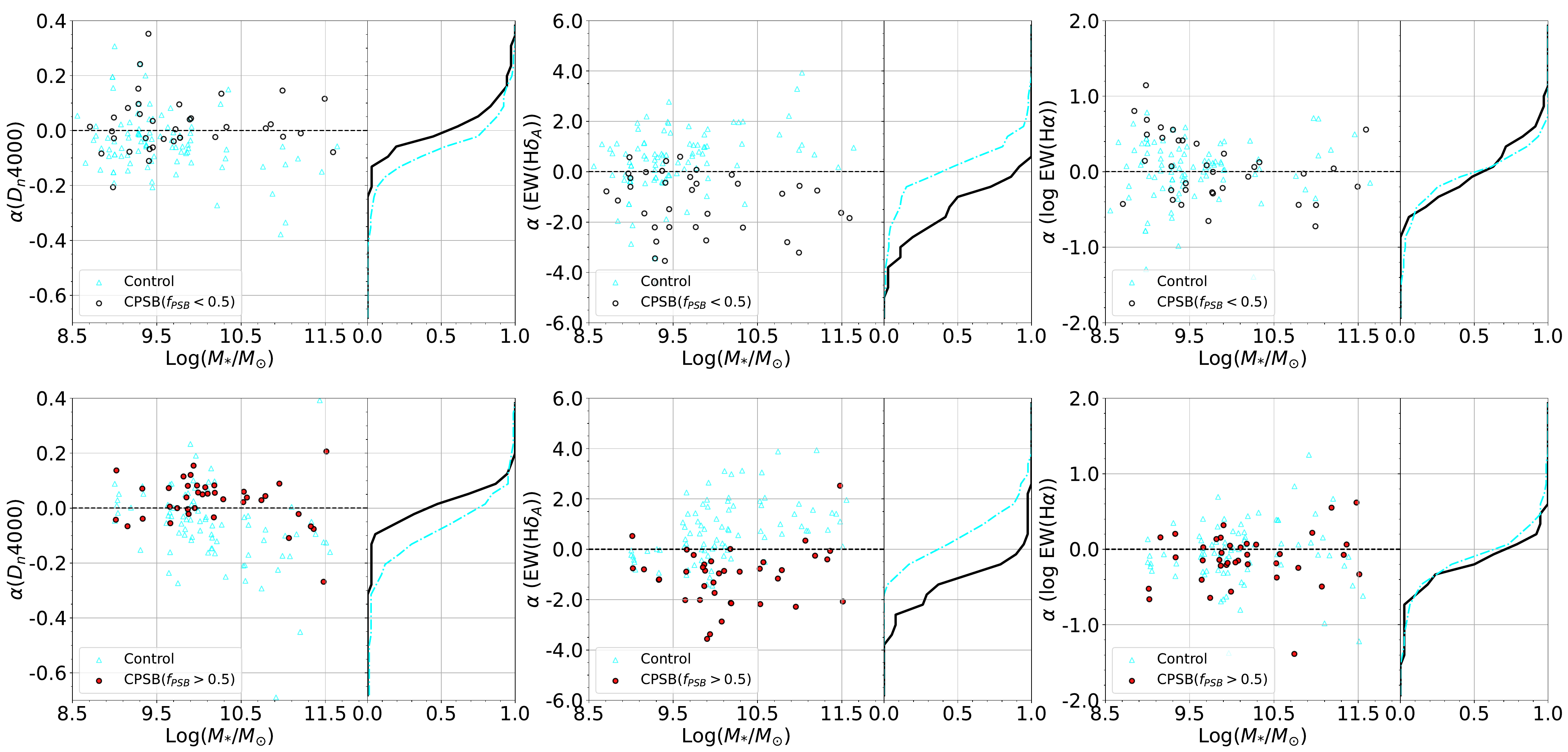}
    \caption{\textcolor{black}{Same as \autoref{fig:alpha}, but for 
    	cPSB galaxies with $f_{\rm PSB}(1.5R_e)<0.5$ (upper panels) and those with 
    	$f_{\rm PSB}(1.5R_e)>0.5$ (lower panel), separately.}}
    \label{fig:fig10}
\end{figure*}

We divide the cPSB sample into two subsamples at $f_{\rm PSB}(1.5R_e) = 0.5$, and 
for each subsample we repeat the analysis of the size-mass relation and recent SFH 
in the same way as above. \autoref{fig:fig9} shows the size-mass relations, and 
\autoref{fig:fig10} shows the slope indices of the three SFH diagnostics as a function of 
stellar mass. Overall, the size-mass relations and recent SFHs of the two subsamples are 
similar to what we found above for the cPSB galaxies as a whole, implying that our main 
results are not dependent on the PSB fraction of the galaxies. It is interesting to note that
the cPSB galaxies with $f_{\rm PSB}(1.5R_e)>0.5$ have higher masses as well as smaller 
sizes at fixed mass when compared to those with $f_{\rm PSB}(1.5R_e)<0.5$. 

\color{black}
\section{Discussion} 
\label{sec:discussion}

\subsection{Implications of the size-mass relation}

A number of previous studies have investigated the structure and/or the 
size-mass relation of PSB galaxies at different redshifts, 
and consistently concluded that the PSB galaxies are more compact relative 
to coeval star-forming and quiescent galaxies 
\citep[e.g.][]{Whitaker2012,Yano2016,Wu2018,Wu2020,Chen2022,Setton2022}.
For instance, using the NEWFIRM Medium-Band Survey (NMBS), 
\citet{Whitaker2012} found a large population of PSB galaxies at $1<z<2$,
which had smaller sizes than older galaxies at those redshifts. 
With the same survey, \citet{Yano2016} examined the structure of 
galaxies at $0.5<z<2$ with different spectral types as classified using 
the UV-to-NIR spectral energy distributions (SEDs), and found the 
PSB galaxies at $z>1.5$ to be significantly smaller than any other types 
of galaxies at the same redshift.
\citet{Wu2018} analyzed the spectroscopically selected PSB galaxies 
at $z\sim0.7$ from the LEGA-C survey, and found these galaxies are 
much smaller than expected from the global relation of normal quiescent galaxies.
A similar result has been recently obtained by \citet{Setton2022} for the 
PSB galaxies from the SQuIGGLE sample, also at $z\sim0.7$, 
which lie $\sim$0.1dex below the size-mass relation of quiescent galaxies.
For the local universe, 
\citet{Chen2022} have recently studied the size-mass relation of the PSB 
galaxies in the SDSS, which are similar to the cPSB galaxies in our 
sample in the sense that the PSB feature is limited to the central 1-2 kpc
of these galaxies. Consistent with their result, our cPSB galaxies at 
intermediate masses ($9.5\lesssim$\lgmstar$\lesssim10.5$) present 
smaller half-light radii than control galaxies of similar mass and SFR.
The smaller size of the PSB galaxies at fixed mass as consistently 
found in previous studies and for different redshifts have been 
usually attributed to mergers or other violent events, which can drive 
gas inward and trigger or enhance the star formation in the inner region 
of galaxies. Obviously, this is consistent with galaxy mergers as the 
origin of PSB galaxies, as discussed in \autoref{sec:intro}.

Thanks to the spatially resolved spectroscopy from MaNGA, 
our sample additionally includes other types of PSB galaxies in which 
the PSB regions are located beyond the galactic center. 
We find the different types of PSB galaxies to show different size-mass relations.
In particular, the smaller size at fixed mass as previously found for 
intermediate-mass galaxies holds only for cPSB galaxies in our sample.
Differently, the iPSB and rPSB closely follow 
the size-mass relation of the control galaxies. 
Our result of the size-mass relation further complicates the evolution 
picture of PSB galaxies, which cannot be purely attributed to galaxy mergers. 
Rather, PSB galaxies as a whole must follow diverse evolutionary pathways,
depending on both stellar mass and the location of the PSB regions.

\subsection{Two broad categories of evolution pathways}


PSB galaxies are widely thought to be in a transition phase between 
star-forming galaxies and quiescent galaxies. The paucity of both 
the ``green-valley'' galaxies and the PSB galaxies imply 
that the time-scale of the transition phase must be very short. 
However, the importance of the PSB phase in galaxy evolution has 
been questioned by some observational studies. For instance, the 
large reservoirs of cold gas detected in many low-$z$ PSB galaxies 
\citep[e.g.][]{Chang2001,Buyle2006,Zwaan2013,French2015}
are in conflict with the gas-poor nature of quenched galaxies, 
and the low incidence of PSB galaxies in some galaxy clusters at
intermediate redshifts is insufficient to explain the quiescent galaxy
population in those clusters \citep[e.g.][]{DeLucia2009,Dressler2013}. 
\citet{Dressler2013} suggested that the PSB galaxies in both 
clusters and the field are already quiescent, but passing through 
a brief PSB phase of the evolutionary cycle within the quiescent
sequence, which could be triggered by tidal interactions or mergers 
with low-mass gas-rich companions. More recently, 
in a comprehensive study of the PSB galaxies in the SDSS, 
\citet{Pawlik2018} found evidence for three different pathways of
galaxies that can lead to a PSB phase, all triggered by violent events
such as galaxy mergers: (1) gas-rich major mergers causing the
transformation of blue-cloud galaxies (via the PSB phase) to the red
sequence, accounting for 60-70\% of the PSB galaxies in their sample,
(2) less violent events causing the cyclic evolution of low-mass 
star-forming galaxies within the blue cloud (the blue-PSB-blue cycle),
and (3) less violent events causing the cyclic evolution of massive
quiescent galaxies within the red sequence (the red-PSB-red cycle). 
The three different pathways of PSB galaxies and their merger origin 
are nicely illustrated by \cite{Pawlik2019} based on the Evolution and 
Assembly of GaLaxies and their Environments (EAGLE) cosmological 
simulation. It is interesting that the same simulation additionally 
presents a fourth pathway of a PSB phase, in which the rapid 
decline in the star formation of a galaxy is caused by environmental 
effects such as ram-pressure stripping, without leading to 
morphological transformation \citep{Pawlik2019}.


Since the PSB galaxies in \citet{Pawlik2018} were selected with the 
SDSS single-fiber spectroscopy, the PSB feature is limited to the central
1-2 kpc of their galaxies, similar to the case of cPSB galaxies in our 
work. Therefore, the three evolutionary pathways suggested by those
authors should be relevant only to the cPSB type. Although the situation 
becomes more complex when the other types are included, 
our results indicate that the PSB galaxies as a whole can be divided 
into two broad categories according to their recent SFH. 
The first category consists of the iPSB-SF and iPSB-Q types, which 
present high similarities to the corresponding control samples in all 
the three SFH diagnostic parameters, likely to follow an inside-out 
quenching process as previously found for the general galaxy population.
The other category consists of the cPSB and rPSB types of galaxies, 
which present younger stellar populations in the center than in the outskirt, 
thus likely to follow an outside-in quenching process. 
The three evolutionary pathways suggested by \citet{Pawlik2018} 
should be applied only to the cPSB type, and so they all belong to the 
second category. The limited size of our sample doesn't allow us to 
distinguish the three pathways of the cPSB type as done in 
\citet{Pawlik2018}, but one can well expect so if a sufficiently large 
sample becomes available considering the (almost) same selection 
criteria of our cPSB galaxies and the SDSS-based PSB galaxies.

\textcolor{black}{We note that, when discussing the quenching pathways 
we haven't considered the quenching of an already quenched but recently 
``rejuvenated'' galaxy (e.g. caused by a merger-induced central starburst), 
which is not the same as the original quenching prior to rejuvenation. 
For a given spaxel, to distinguish these two quenching processes would 
need careful stellar population synthesis modeling. For a galaxy in which 
the onset of quenching varies radially as indicated by radial gradients of 
the recent SFH in relatively massive galaxies \citep[e.g.][]{Wang2018}, 
the original quenching and the quenching after rejuvenation can be distinguished 
according to the radial variation of the three SFH diagnostic parameters. 
For instance, we can certainly rule out the case of quenching after a central 
starburst for most of our galaxies. In this case, one would expect small 
\dbalmer\ in the central region due to the recent starburst as well as large 
\dbalmer\ in the originally quenched outerskirt. As can be seen from the 
first row of \autoref{fig:cen_re}, however, we find almost no spaxels to be 
located in the upper-left corner of the \dbalmer\ diagram for all the PSB types. 
All the galaxies are found at either the lower-left corner 
with relatively low \dbalmer\  or the upper-right corner (below the 1:1 relation) 
with large \dbalmer, in both the center and the outskirt. This fact suggests 
the original quenching to be the dominant quenching process for the galaxies 
in our sample. One exception may happen if the central starburst occurs very 
recently. In this case, the \dbalmer\ in the central region doesn't have time to 
substantially decrease, and as a result the galaxy would have similarly large 
\dbalmer\ at both the center and outskirt and meanwhile have large central 
\ewha\ and weak \ewha\ at the outerskirt. Such cases are not observed in 
\autoref{fig:cen_re}, too. This imply that, it rarely happens that a fully-quenched 
galaxy becomes rejuvenated and restarts quenching only in their central region; 
rather, it is more likely that star formation rejuvenation happens over a wide range 
of galactic radii (although the central region may be more enhanced in star formation 
due to the merger-driven gas inflow),  and the consequent quenching happens 
earlier at larger radii and propagates inward (thus still in an outside-in manner).
In any case, the outside-in quenching inferred from our data for the cPSB and 
rPSB galaxies cannot be a consequence of a central starburst.}

\subsection{The evolutionary relation between rPSB and cPSB types}

\begin{figure*}[htbp]
	\centering 
	\includegraphics[width=\textwidth]{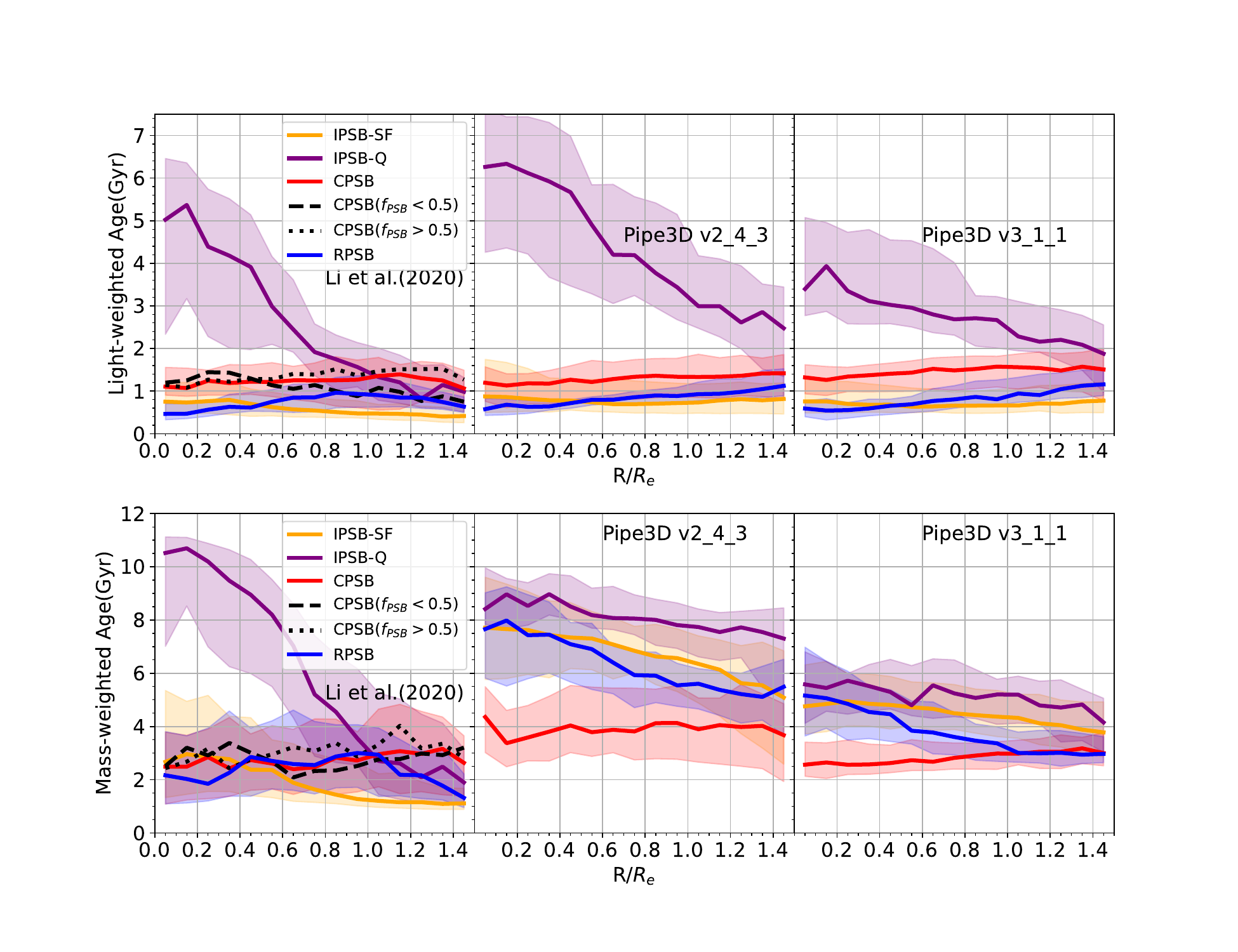}
	\caption{$Top$ $row$: radial gradients of the average lighted-weighted age of PSB samples. The shadowed regions show the 30\% to 70\% percentile of the distributions. From left to right, the spectral fitting codes are different. \textcolor{black}{The average results of cPSB types with different $f_{\rm PSB}$ are highlighted by dashed and dotted lines.} $Bottom$ $row$: same as the top row, but showing the mass-weighted age result.}
	\label{fig:stellar_age}
\end{figure*}

The rPSB and cPSB types are similar in \dbalmer~and~\ewhd, but 
different in \ewha, with the rPSB galaxies showing larger \ewha\ 
in their centers. As a result, the rPSB galaxies mostly fall in the 
star-forming sequence and the blue cloud, while the cPSB galaxies 
mostly fall below the star-forming sequence and in the green valley.
These similarities and differences appear to suggest that the 
rPSB and cPSB may have the same origin (e.g. mergers or environmental effects) but at different evolutionary stages. In the outside-in quenching 
process as inferred for both types, the cessation of the star formation 
in a galaxy in this category will first happen in its outskirt, and 
then extends to smaller radii. During this process, the galaxy  
will be observed as a rPSB galaxy at an early stage, and then a 
cPSB galaxy later on. 

The two types also differ in the size-mass relation. The rPSB 
type follows the same relation as the control sample and the 
average relation of star-forming galaxies, while the cPSB galaxies 
have smaller half-light radii at intermediate-to-low masses. 
This result is also consistent with the rPSB galaxies being 
progenitors of cPSB galaxies. In a recent study of PSB galaxies 
at $z\sim0.7$, \citet{Setton2022} studied the evolution of the 
PSB galaxy properties as a function of the time since the star 
formation was quenched. The sizes of PSB galaxies were found 
to show a negative or no correlation with the time since quenching. 
If the outside-in process is taken into account, one would 
expect a PSB galaxy to become smaller or similarly sized when 
it evolves to a later stage as a cPSB galaxy. 
This conjecture can be tested using current hydrodynamic 
simulations of galaxy formation. 

We note that, in an earlier study of the MaNGA-based PSB galaxy 
sample, \citet{Chen2019} found the rPSB and cPSB galaxies
are quite different in the radial gradient in mass-weighted stellar 
age, based on data products produced by the spectral fitting code 
{\tt Pipe3D} \citep{San2016}. This result indicated that cPSB
and rPSB are not simply different evolutionary stages of the 
same event. Instead of using stellar population properties derived 
from the technique of stellar population synthesis such as {\tt Pipe3D}, 
in our work we opt for the three SFH indicators which are directly 
measured from the stacked spectrum of each region and thus 
are model-independent. In fact, the spectral fitting 
code used in this work also provides measurements of stellar 
population properties. In~\autoref{fig:stellar_age},
we compare the radial gradients in both light-weighted (upper panels)
and mass-weighted (lower panels) age for the four types of PSB samples, 
as obtained by applying our code (the left panels) and the {\tt Pipe3D} 
(version 2.4.3; the middle panels) 
which is the same version as used in \citet{Chen2019}. 
In addition, in the rightmost panels, we show the results based 
on the new version of {\tt Pipe3D} (version 3.1.1) which has 
replaced the stellar library of MILES by the recently available 
library of MaStar \citep{Yan2019}. As can be seen, the different 
codes produce similar results for light-weighted age, but the 
mass-weighted profiles are quite different, particularly for 
the rPSB sample (the blue lines) and the iPSB-SF sample (the yellow lines).
The strong (negative) gradients in mass-weighted age for 
the rPSB type and the flat profile for the cPSB as reported 
in \citet{Chen2019} are also seen in our figure (the middle lower panel).
However, the gradient in the rPSB galaxies becomes rather weak
when the new version of {\tt Pipe3D} is applied, a result which is 
close to (though not exactly the same as) the result of our code
where we see little difference between the cPSB and rPSB samples.
The discrepancies between the different codes reflect the known 
issue of stellar population modeling which suffers from 
strong degeneracies between model parameters such as 
age, metallicity, dust attenuation, etc. 

\textcolor{black}{In the left-hand panels of \autoref{fig:stellar_age},
	we also show the stellar age profiles for the two subsamples of 
	cPSB galaxies, with $f_{\rm PSB}<0.5$ and $f_{\rm PSB}>0.5$ 
	as analyzed in \autoref{subsec:fpsb}. It is seen that the two 
	subsamples show very similar profiles, implying that the cPSB 
	galaxies of different $f_{\rm PSB}$ fall in the same category 
	as mentioned above. The broad range of $f_{\rm PSB}$ found for 
	the cPSB galaxies may reflect the galaxy-to-galaxy variation of 
	the relative timescales between the PSB phase and the overall 
	quenching procedure of the host galaxy. If so, the quenching timescale
	must be somehow dependent on the stellar mass of galaxies, 
	as indicated by the correlation of $f_{\rm PSB}$ with \mstar\ 
	(see \autoref{subsec:fpsb}).}

\citet{Otter2022} have recently analyzed ALMA observations of the
$^{12}$CO(0-1) distribution in 13 cPSB and rPSB galaxies selected 
from the MaNGA survey. Their results show stellar disturbances 
as well as centrally concentrated and highly disturbed molecular 
gas in both cPSB and rPSB galaxies, indicating a common merger origin. 
If the two types of galaxies have distinct origins and are not sequential 
evolutionary phases, as pointed out by \citet{Otter2022}, a natural 
question would arise: why do some mergers lead to a cPSB phase 
whereas others lead to a rPSB phase? The authors suggested that
outflows could be an alternative process primarily impacting 
cPSB galaxies, considering the fact that this type of galaxies are 
more disturbed in both ionized and molecular gas when compared 
to rPSB galaxies. Using integral-field spectroscopy 
from the SAMI Galaxy Survey, \cite{Owers2019} found that
the PSB galaxies in galaxy clusters mostly present PSB regions in 
their outskirt, and concluded that these galaxies follow an outside-in 
quenching driven by ram pressure stripping.

In summary, the rPSB type of galaxies appear to have multiple
origins depending on environment. The majority of rPSB galaxies, which are 
located in the field and loose groups with dynamical conditions 
more favorable for galaxy interactions and mergers, are likely
to have the same merger origin and experience the same sequential 
evolutionary pathway as the cPSB type. For those located in dense 
environments such as galaxy clusters, ram-pressure stripping may play 
a more important role. Both mergers and ram-pressure stripping lead 
to outside-in quenching.

\subsection{The evolutionary relation between iPSB-SF and iPSB-Q types}

The majority of the PSB galaxies in our sample are classified as iPSB, 
either iPSB-SF or iPSB-Q according to their global SFR, and these two 
types behave similarly in many aspects including the size-mass relation 
and the SFH indicators. They differ mainly 
by the global SFR, by definition, with one in the star-forming sequence and one 
in the quiescent sequence. Thus, it is natural to speculate that they are 
different evolutionary stages of the same evolutionary pathway, with the iPSB-SF 
type being progenitors of the iPSB-Q type. However, we would like to caution 
that this conjecture may be true only for a portion of the iPSB galaxies; for 
some galaxies, the iPSB-SF and iPSB-Q types may be induced by distinct 
events. For instance, ``rejuvenated'' star 
formation in the outskirt of an already quiescent galaxy due to recent gas 
accretion can lead to an iPSB phase, after which the galaxy returns to the 
quiescent sequence. This is similar to the red-PSB-red cycle as proposed 
in previous studies \citep[e.g.][]{Dressler2013,Pawlik2018}. 
More detailed studies of the stellar populations and environment for these 
types of galaxies, both observational and theoretical,  would be needed in 
future when larger samples and more reliable techniques of stellar population 
synthesis become available. 

\section{Summary} 
\label{sec:summary}

In this work, we use the spatially resolved spectroscopy from the final data 
release of the MaNGA survey to identify post-starburst galaxies (PSB) 
in the local Universe. The PSB galaxies are identified in a two-step method. 
For each galaxy from MaNGA,  we first select regions with a size of $\sim$kpc
and strong H$\delta$ absorption, from which we 
then select PSB regions that have relatively weak H$\alpha$ emission. 
For each region, we have stacked the original spectra of its spaxels to 
obtain a spectrum of high S/N, and we perform full spectral fitting to the 
stacked spectrum to measure emission line parameters and stellar indices 
that are used for the identification and scientific analysis in this work. 
Out of the 10,010 galaxies from MaNGA, we have identified a sample of 
489 PSB galaxies, each with at least one PSB region. These galaxies  are 
then classified into three different types according to the location of their 
PSB regions: 94 cPSB galaxies in which the PSB regions 
are found at the galactic center, 85 rPSB galaxies in which the PSB regions 
are off-center and form a ring-like shape, and 310 iPSB galaxies in which 
the PSB regions present irregular locations and shapes. 
Each of the iPSB galaxies is further classified as either iPSB-SF if located 
in the star-forming sequence in the diagram of star formation rate (SFR)
versus stellar mass, or iPSB-Q if located in the quiescent sequence in 
the same diagram. 

We first compare the size-mass relation of the four types of PSB galaxies
with control samples of normal galaxies that are closely matched in redshift, 
stellar mass and SFR. We then study the recent star formation history (SFH)
of the PSB samples and the corresponding control samples, using three 
spectral indices to indicate the recent SFH: \ewha~(equivalent width 
of the H$\alpha$ emission line), \ewhd~(equivalent width of the H$\delta$ 
absorption line), and \dbalmer~(depth of the break at 4000\AA). 
\textcolor{black}{In addition, we examine the potential dependence on the 
	fraction of PSB regions in our galaxies.}

Our conclusions can be summarized as follows. 

\begin{itemize} 
  \item Different types of PSB galaxies show different size-mass relations:
  iPSB-SF galaxies closely follow the size-mass relation of 
  normal star-forming galaxies, iPSB-Q galaxies follow the relation of 
  quiescent galaxies, rPSB galaxies appear to fall in between, and 
  cPSB galaxies follow the relation of star-forming galaxies at both 
  high and low masses and that of quiescent galaxies at intermediate masses. 
  \item When compared to control galaxies of similar SFR, redshift, and mass, 
  the cPSB galaxies have smaller sizes at 
  $9.5\lesssim \log_{10}(\rm M_\ast/M_\odot)\lesssim 10.5$, 
  while the other types of PSB galaxies show similar sizes 
  to their control galaxies at a given mass.
  \item The iPSB-SF sample shows no/weak gradients in 
  \dbalmer, \ewhd~and~\ewha, consistent with the global 
  star-forming status of this type of galaxies.
  \item The iPSB-Q sample shows negative gradients in \dbalmer, 
  positive gradients in \ewhd, and no gradients in \ewha. 
  This result indicates older stellar populations in the inner regions, 
  consistent with an ``inside-out'' quenching process.
  \item The cPSB sample shows positive 
  gradients in \dbalmer, negative gradients in \ewhd, and no gradients 
  in \ewha, indicating younger stellar populations in the inner regions
  and an ``outside-in'' quenching process.
  \item The rPSB sample is similar to the cPSB sample in 
  the gradients in both \dbalmer~and~\ewhd, but shows negative 
  gradients in \ewha. This result implies that rPSB and cPSB 
  galaxies are different stages of the same evolutionary pathway.
  \textcolor{black}{\item Both rPSB and iPSB galaxies have relatively small 
  fractions of PSB regions, with $f_{\rm PSB}<0.5$ within 1.5$R_e$, 
  while cPSB galaxies spand a full range in $f_{\rm PSB}$. The size-mass 
  relation and recent SFHs of cPSB galaxies show no dependence on 
  $f_{\rm PSB}$.}
  \item The four types of PSB galaxies can be broadly divided into 
 two categories in terms of evolutionary pathway: (1) iPSB-SF 
 and iPSB-Q in the first category which have similar SFH to normal 
 galaxies and appear to prefer an inside-out quenching process,
 and (2) cPSB and rPSB in the other category which prefer 
 an outside-in quenching process, likely driven by disruption events
 such as mergers which make the galaxies more compact than 
 the average quiescent galaxies.
\end{itemize}

\section*{Acknowledgments}
\textcolor{black}{We're grateful to the anonymous referee for the helpful comments on our paper.}
This work is supported by the National Key R\&D Program of China
(grant No. 2022YFA1602902), and the National Natural Science 
Foundation of China (grant Nos. 11821303, 11733002, 11973030, 
11673015, 11733004, 11761131004, 11761141012). 

Funding for the Sloan Digital Sky 
Survey IV has been provided by the 
Alfred P. Sloan Foundation, the U.S. 
Department of Energy Office of 
Science, and the Participating 
Institutions. 

SDSS-IV acknowledges support and 
resources from the Center for High 
Performance Computing  at the 
University of Utah. The SDSS 
website is www.sdss.org.

SDSS-IV is managed by the 
Astrophysical Research Consortium 
for the Participating Institutions 
of the SDSS Collaboration including 
the Brazilian Participation Group, 
the Carnegie Institution for Science, 
Carnegie Mellon University, Center for 
Astrophysics | Harvard \& 
Smithsonian, the Chilean Participation 
Group, the French Participation Group, 
Instituto de Astrof\'isica de 
Canarias, The Johns Hopkins 
University, Kavli Institute for the 
Physics and Mathematics of the 
Universe (IPMU) / University of 
Tokyo, the Korean Participation Group, 
Lawrence Berkeley National Laboratory, 
Leibniz Institut f\"ur Astrophysik 
Potsdam (AIP),  Max-Planck-Institut 
f\"ur Astronomie (MPIA Heidelberg), 
Max-Planck-Institut f\"ur 
Astrophysik (MPA Garching), 
Max-Planck-Institut f\"ur 
Extraterrestrische Physik (MPE), 
National Astronomical Observatories of 
China, New Mexico State University, 
New York University, University of 
Notre Dame, Observat\'ario 
Nacional / MCTI, The Ohio State 
University, Pennsylvania State 
University, Shanghai 
Astronomical Observatory, United 
Kingdom Participation Group, 
Universidad Nacional Aut\'onoma 
de M\'exico, University of Arizona, 
University of Colorado Boulder, 
University of Oxford, University of 
Portsmouth, University of Utah, 
University of Virginia, University 
of Washington, University of 
Wisconsin, Vanderbilt University, 
and Yale University.

\bibliographystyle{aasjournal}
\bibliography{sample631}{}

\end{document}